\documentclass[twocolumn,twoside]{ncnsd}
\usepackage{amsmath}

\def\BibTeX{{\rm B\kern-.05em{\sc i\kern-.025em b}\kern-.08em
    T\kern-.1667em\lower.7ex\hbox{E}\kern-.125emX}}

\begin{document}

\title{Bose-Einstein Condensation Dynamics from the Numerical Solution
of the Gross-Pitaevskii Equation by the Pseudospectral Method}

\author{Paulsamy Muruganandam and Sadhan K. Adhikari
\thanks{P.~Muruganandam (corresponding author) is with the Centre for
Nonlinear Dynamics, Department of Physics, Bharathidasan
University, Tiruchirapalli - 620 024, India, e-mail:
anand@cnld.bdu.ac.in. and S.~K.~Adhikari is with Instituto de
F\'{\i}sica Te\'orica, Universidade Estadual Paulista, S\~ao
Paulo, 01405-900 SP Brasil, e-mail: adhikari@ift.unesp.br.}}

\markboth{NATIONAL CONFERENCE ON NONLINEAR SYSTEMS \& DYNAMICS}
{INDIAN INSTITUTE OF TECHNOLOGY, KHARAGPUR 721302, DECEMBER 28-30,
2003}

\maketitle

\begin{abstract}
We study certain stationary and time-evolution problems of trapped
Bose-Einstein condensates of weakly interacting alkali atoms
described by a nonlinear Gross-Pitaevskii (GP) equation. We
suggest a pseudospectral method involving Laguerre polynomials to
solve the time dependent GP for a spherically symmetric trap
potential. The radial wavefunction and energy values have been
calculated for different nonlinearities. Further, we study the
effect of suddenly changing the interatomic scattering length or
harmonic oscillator trap potential in the condensate. We also
investigate the frequency of oscillation due to the variation in
the strength of nonlinearity.
\end{abstract}


\section{Introduction}

\PARstart{T}{he} experimental realization\cite{Dalfovo_etal} of
Bose-Einstein condensates (BECs) in dilute weakly-interacting
trapped bosonic atoms at ultra-low temperature initiated intense
theoretical effort to describe the properties of the condensate.
The properties of a condensate at zero temperature are usually
described by the time-dependent, nonlinear, mean-field
Gross-Pitaevskii (GP) equation\cite{Dalfovo_etal}. The effect of
the interatomic interaction (few-body correlation) leads to a
nonlinear term in the GP equation which complicates the solution
procedure. Although there have been previous studies of the
solution of the GP equation for stationary or time-independent
problems, virtually all time-dependent studies in realistic cases,
e.g., in three space dimensions, have employed approximate
approaches rather than exact numerical solution of the GP
equation.

A numerical study of the time-dependent GP equation is of
interest, as this can provide solution to many stationary and
time-evolution problems. The time-independent GP equation yields
only the solution of stationary problems. As our principal
interest is in time evolution problems, we shall only consider the
time-dependent GP equation in this paper.

Here we propose a pseudospectral time-iteration method for the
solution of the three-dimensional GP equation with spherically
symmetric trap. In the pseudospectral method the unknown wave
function is expanded in terms interpolating polynomials
\cite{Weideman_Reddy}. When this expansion is substituted into
the GP equation, the (space) differential operators operate on a
set of known polynomials and generate a differentiation matrix
operating on the unknown coefficients. Consequently, the
time-dependent partial-differential nonlinear GP equation in
space and time variables is reduced to a set of coupled ordinary
differential equations (ODEs) in time which is solved by a
fourth-order adaptive step-size controlled Runge-Kutta method
\cite{Press_etal} using successive time iteration. Reinhardt and Clark 
\cite{Reinhardt_Clark} have employed pseudospectral methods
for the solution of the GP equation, where a variable step forth
order Runge-Kutta time propagator was used, as in the present
work. A pseudospectral Fourier-sine basis was used in 
\cite{Reinhardt_Clark} for finite traps, and a corresponding 
complex pseudospectral basis was used for systems with periodic 
boundary conditions.

Such a pseudospectral method is used in \cite{ska_pm_jpb:2} with
Hermite polynomials as the interpolant for the case of completely
anisotropic trap potential. However, when the system has radial or
axial symmetry the three dimensional problem reduces to one or two
dimensions. In such situations, it will be more advantageous to
employ the integration in reduced dimensions as it saves
considerable amount of computer memory and time. In this paper we
considered the GP equation for spherically symmetric trap
potential.

In section \ref{gpe} we describe briefly the three-dimensional,
time-dependent GP equation with spherically symmetric  trap. The
pseudospectral method is described in section \ref{psrk}. In section
\ref{numeric} we report the numerical results for the wave
function and energy for different nonlinearities as well as  an
account of our study on the effect of sudden changes in the
interatomic scattering length or harmonic oscillator trap
potential in the condensate. An analysis on the frequency of
oscillation with respect to variation of the strength of
nonlinearity is also reported. Finally, in section
\ref{conclude} we present our conclusions.

\section{Nonlinear Gross-Pitaevskii Equation}\label{gpe}
At zero temperature, the time-dependent Bose-Einstein
condensate wave function $\Psi({\bf r};\tau)$ at position ${\bf
r}$ and time $\tau $ may be described by the following  mean-field
nonlinear GP equation \cite{Dalfovo_etal}
\begin{align}
\left[ -\frac{\hbar^2\nabla ^2}{2m}
+ V({\bf r})
+ gN_0|\Psi({\bf
r};\tau)|^2
-i\hbar\frac{\partial
}{\partial \tau} \right]\Psi({\bf r};\tau)=0. \label{a}
\end{align}
Here $m$ is the mass and  $N_0$ the number of atoms in the
condensate, $g=4\pi\hbar^2 a/m $ the strength of interatomic
interaction, with $a$ the atomic scattering length. The
normalization condition of the wave function is $ \int d{\bf r}
|\Psi({\bf r};\tau)|^2 = 1.$

The three-dimensional trap potential is given by $  V({\bf r})
=\frac{1}{2}m (\omega_x^2\bar  x^2+\omega_y^2\bar  y^2+\omega_
z^2\bar z^2)$, where $\omega_x \equiv \omega_0$, $\omega_y$, and
$\omega_z$ are the angular frequencies in the  $x$, $y$ and $z$
directions, respectively, and   ${\bf r}\equiv (\bar x,\bar y,\bar
z)$ is the radial vector.

\subsection{Spherically Symmetric Case}

In the spherically symmetric trap, i.e., $\omega_x = \omega_y =
\omega_z \equiv \omega_0$ the trap potential is given by $V({\bf
r})=\frac{1}{2}m\omega_0 ^2\tilde{r}^2$, where $\omega_0$ is the
angular frequency and $\tilde{r}$ the radial distance. The wave
function can be written as $\Psi({\bf r};\tau)=\psi(\tilde
r,\tau)$. After a transformation of variables to dimensionless
quantities defined by $r =\sqrt 2\tilde r/l$, $t=\tau \omega_0$,
$l\equiv \sqrt {(\hbar/m\omega_0)} $ and
$\phi(r,t)\equiv\varphi(r,t)/r = \psi(\tilde r,\tau)(4\pi
l^3)^{1/2}$, the GP equation in this case becomes
\begin{eqnarray}
\left[-\frac{\partial^2}
{\partial r^2}+\frac{r^2}{4}+{\cal N}\left|
\frac{\varphi(r,t)}{r}
\right| ^2 -i\frac{\partial }{\partial t}\right] \varphi (r,t)=0,
\label{c}
\end{eqnarray}
where ${\cal N}=N_0a/l$. The normalization condition for the wave
function is
\begin{eqnarray}\label{n1}
\int_0^{\infty} \mid \varphi(r,t) \mid^2\,dr = 2 \sqrt{2}.
\end{eqnarray}

\section{Pseudospectral Runge-Kutta (PSRK) method}
\label{psrk}

The main idea of this method is that the spatial differentiation
operators in equation (\ref{c}) are replaced by appropriate
differential matrices which are derived from spectral or
pseudospectral collocation. Then the time integration is carried
out by any ordinary differential equation (ODE) solver.

In the pseudospectral approximation, an unknown function $f(x)$ is
expanded in terms of weighted interpolants of the form
\cite{Weideman_Reddy,Fornberg:1996}
\begin{equation}
f(x) \approx p_{n}(x) = \sum_{j=0}^{n}
\frac{\alpha(x)}{\alpha(x_j)}\xi_{j}(x)f_{j}.\label{eq:fx}
\end{equation}
Here $\{ {x_j}\} _{j=0}^{n}$ is a set of distinct interpolation
nodes,  $f_j = f(x_j)$, $\alpha(x)$ is a weight function, the
interpolating functions $\{ \xi_j(x)\}^{n}_{j=0}$ satisfy
$\xi_j(x_k)=\delta_{jk}$ (the Kronecker delta) and involve
orthogonal polynomials of degree $(n)$, so that
$f(x_k)=p_{n}(x_k)$, $k=0,1,\ldots,n$. One could use orthogonal
polynomial such as, Chebyshev, Hermite, Laguerre and Legendre as
the interpolant. Even non-polynomial interpolants like Fourier
(spectral) expansion of the function in terms of periodic cosine
and sine functions can be used in the case of periodic boundary.
The choice of interpolant depends on the domain and nature of
boundary conditions.

Differentiating (\ref{eq:fx}) $m$ times, the $m^{\mbox{th}}$
derivative of $f(x)$ at the nodes is given by,
\begin{equation}
f^{(m)}(x) \approx \sum_{j=0}^{n}\frac{d^m}{dx^m}\left[
\frac{\alpha(x)}{\alpha(x_j)}\xi_{j}(x) \right]_{x=x_k}f_j,
\end{equation}
with $k=1,\ldots,n$. Here the derivative operator is represented
by $D^{(m)}$ which is the \emph{differentiation matrix} with
elements,
\begin{equation}
D^{(m)}_{k,j} = \frac {d^m}{dx^m} \left[  \frac{\alpha(x)}{\alpha(x_j)}
\xi_j(x) \right] _{x=x_k}.\label{diff_mat}
\end{equation}
Thus, the numerical differentiation of the function $f(x)$,
$f^{(m)}=D^{(m)}f$ is performed as the  matrix-vector product
where $f$ and $f^{(m)}$ are the vectors of the function and its
derivative values evaluated at the nodes.

In the present work we consider Laguerre polynomials $L_n(x)$ as
the interpolating functions with the weight function $\alpha(x) =
\mbox{e}^{-\frac{1}{2}bx}p(x)$. The reason behind the choice of
Laguerre polynomials is that they are well defined in the interval
$x\in [0,\infty)$ and satisfy the boundary conditions of the wave
function of the GP equation~\cite{Edwards_Burnett}. The choice
$x=r^2$  gives exact eigenfunctions of the GP equation (\ref{c})
when the nonlinearity term ${\cal N}=0$. Actually, the associated
Laguerre polynomials are more appropriate as they are the radial
eigenfunctions of linearized GP equations. However, from a
numerical point of view the use of both of the polynomials yield
more or less the same result. Also it has not been known whether
there is a significant advantage of improvement in the results by
considering associated Laguerre polynomials\cite{Weideman_Reddy}.

Consequently, the partial differential equation (\ref{c}) is
reduced to a set of coupled ODEs in the time variable $t$
involving $\xi_j, j =0,1,\ldots,n$. In this way we obtain a set of
ODEs by \emph{collocating} the original equations on a suitable
set of points (the roots of some Laguerre polynomials).

For solving the set of ODEs we use the adaptive step-size control
based on the embedded Runge-Kutta formulas due to Fehlberg
\cite{Press_etal}, which gives a definite clue about how to modify
the step size in order to achieve a desired accuracy in a
controlled way. For orders $M$ higher than four of the Runge-Kutta
formula, evaluation of more than $M$ functions (though never more
than $M+2$) is required. This makes the classic fourth order
method requiring the evaluation of four functions  more economic.
Fehlberg suggested a fourth-order and a fifth-order method each
requiring the evaluation of six functions. The difference between
the results of these two gives the error $\delta$ in the
fourth-order method with a step size $h$, where $\delta$ scales as
$h^5:$ $\delta \propto h^5$. This scaling immediately gives the
factor by which the step size $h$ should be reduced, so that a
desired $\delta$ can be obtained. The detailed fourth-order and
fifth-order Runge-Kutta formulas of Fehlberg are given in
\cite{Press_etal}. We use these formulas with the constants given
by Cash and Karp also tabulated in \cite{Press_etal}. For the
present problem we find that the use of Cash-Karp constants  in
the  Fehlberg formulas leads to more accurate results than the
original constants due to Fehlberg.

Using the differentiation matrix (\ref{diff_mat}), the GP equation
is discretized. The grid points are the roots of the Laguerre
polynomial $L_{n}(x_j)=0$. However, the actual $x_j$ values
employed are obtained by scaling these roots by a constant factor
so that most of the roots fall in the region where the condensate
wave function is sizable and only a few points are located in the
region where the wave function is negligible. In the present
paper, all the numerical simulations have been carried out with
$n=31$ grid points.

\section{Numerical simulation}
\label{numeric}

\subsection{Wavefunction and energy}

Once the GP equation (\ref{c}) is reduced to a set of coupled
ODEs, the time integration is carried out with an initial seed
wavefunction. We use the solution corresponding to the linearized
GP equation as the seed. The final wavefunction is obtained by
adding the nonlinear terms in small steps until the desired
nonlinearity is achieved. Then the stationary wave function $\phi
$ and the parametric energy $\mu$ (the chemical potential) can be
extracted from the evolution of the time-dependent GP equation
over a macroscopic interval of time
\cite{Ruprecht_etal,ska_pre_63_056704}.

First, we are interested to calculate the wavefunction and energy
with different nonlinearity strengths. For this purpose, we
numerically integrate equation (\ref{c}) with the following
normalized analytic solution,
\begin{equation}
\varphi(r)=\left(\frac{2}{\pi^{\frac{1}{4}}}\right)r\,
\exp\left(-\frac{r^2}{4}\right).
\label{eq_ini}
\end{equation}
The above $\varphi(r)$ corresponds to the ground state of
(\ref{c}) for spherical symmetry with the coefficient of the
nonlinear term set to zero. During the integration the coefficient
of the nonlinear term is increased from $0$ at each step by
$\delta_n=0.001$ until the final value of nonlinearity ${\cal N}$
is attained at a time called time $t=0$. This corresponds to the
final solution and this solution is found to be stable. The norm
of the wavefunction  is conserved during the integration. However,
it is of advantage to reinforce numerically the proper
normalization of the wave function given by (\ref{n1}) after
finite number of RK steps in order to improve the precision of the
result. As the grid points are the corresponding Laguerre roots,
the integral in (\ref{n1}) can easily be evaluated by appropriate
Gauss-Laguerre formula.
\begin{figure}[!ht]
\centering\includegraphics[width=0.95\columnwidth]{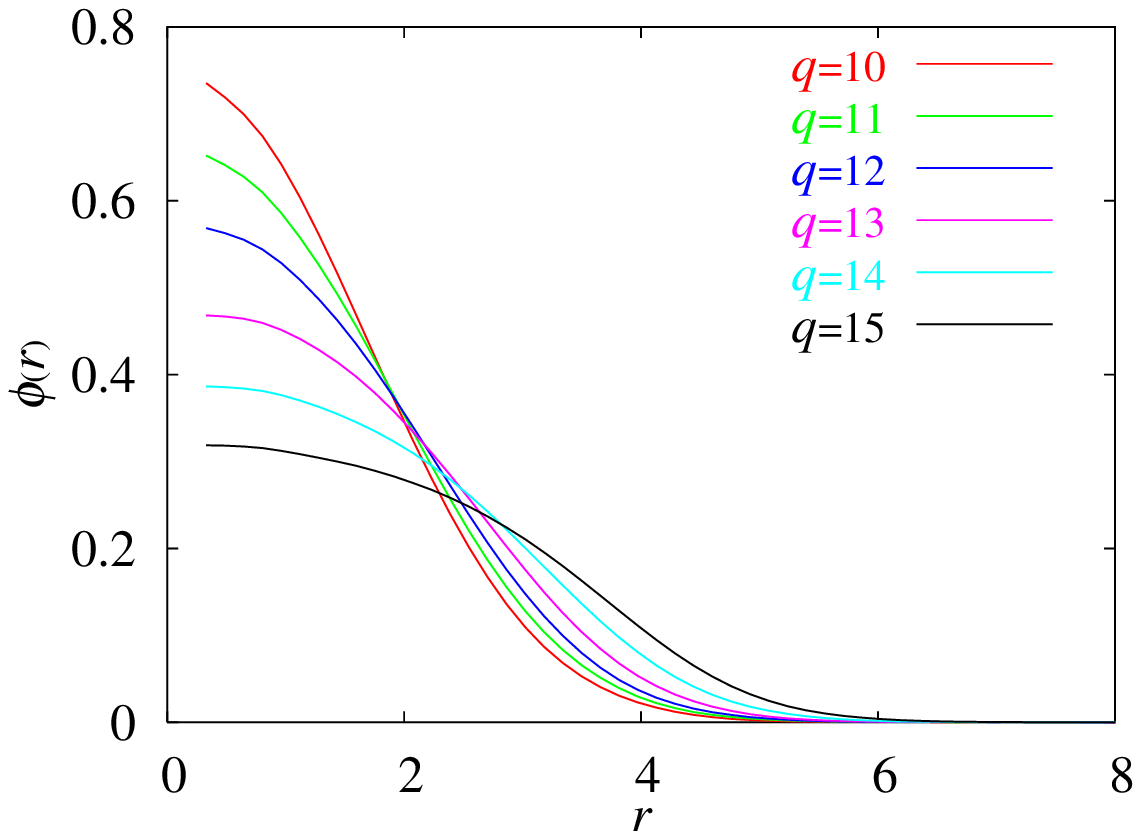}
\caption{Figure showing the radial wavefunctions of the condensate for
different number of atoms $N_0 = 2^q$.}
\label{f1}
\end{figure}
Figure~\ref{f1} depicts the radial wavefunction of the condensate for different
nonlinearity strengths.
\begin{table}[!ht]
\begin{center}
\caption{The chemical potential $\mu$ for various number of condensate atoms.}
\label{table1}
\begin{tabular}{rrl|rrl}
\hline
\multicolumn{1}{c}{$N_0$} & \multicolumn{1}{c}{$\cal N$}&
\multicolumn{1}{c|}{$\mu$} & \multicolumn{1}{c}{$N_0$} &
\multicolumn{1}{c}{$\cal N$} & \multicolumn{1}{c}{$\mu$} \\
\hline
1024  &  0.50 & 1.824 & 2048 &  1.00 & 2.065 \\
4096  &  1.98 & 2.429 & 8192 &  3.97 & 2.969 \\
16384 &  7.93 & 3.722 & 32768 & 15.86 & 5.748 \\
65536 & 31.72 & 6.155 & 131072 & 63.44 & 8.041 \\
\hline
\end{tabular}
\end{center}
\end{table}
The stationary wavefunction has trivial time dependence $\phi(r,t)
= \phi(r)\exp(-i\mu t)$. Thus, the parametric energy $\mu$
(chemical potential) can be extracted from the evolution of the GP
equation over a macroscopic interval of time. In
Table~\ref{table1}, we present the chemical potential of several
BECs with different number of atoms for the spherically symmetric
trap potential. The typical value of the harmonic trap frequency
is $\omega_0 = 87\,\mbox{rad}/s$ and scattering length is
$a=52a_0$ (for Na) where $a$ is the Bohr
radius\cite{Dalfovo_etal}. The energy values calculated here for
different nonlinearity strengths are compared reasonably well with
the results of Schneider and Feder~\cite{Schneider_Feder} by
time-independent approach.

\subsection{Time evolution}

Next we consider certain time-evolution problems which can be
tackled more effectively with the present pseudospectral method.
From the experimental point of view, it is of great interest to
study the effects of changing (i) the trap potential and (ii) the
atomic scattering length. The former can be achieved in the
laboratory by changing the magnetic field responsible for trapping
while it has been possible to modify the scattering length in a
controlled fashion by exploiting a \emph{Feshbach resonance}
\cite{Inouye_etal}. In this paper we restrict ourselves to the
study on the effect of suddenly changing both the trap potential
as well as the atomic scattering length.

\subsubsection{Effect of changing trap potential and scattering length}

As said above, one can increase or reduce suddenly the strength of
the nonlinear term through a change of the scattering length in
this fashion and study the oscillation of the condensate
thereafter. In both these cases one observes the time evolution of
the root mean square (rms) radius $\langle r \rangle_{\mbox{rms}}$
of the condensate. For the time evolution study we consider a
previously formed condensate with finite number of atoms, for
example, $N_0 = 16384$ prepared at $t=0$ as in the study of the
stationary problem. We then inflict the four following changes in
the system. At $t=0$ we (a) increase or (b) decrease suddenly the
coefficient of the harmonic oscillator $r^2/4$ term in (\ref{c})
by a factor of 2. Next at $t=0$ we (c) increase or (d) decrease
suddenly the coefficient of the nonlinear term ${\cal N}$ in
(\ref{c}) by a factor of 2.

When the harmonic oscillator term is doubled or the nonlinearity
halved, the system is compressed and the rms radius oscillates
between its initial value and a smaller final value.  When the
harmonic oscillator term is halved or the nonlinearity doubled,
the system expands and the rms radius oscillates between its
initial value and a larger final value.

We have calculated the frequency of oscillation of the rms radius
in all the four cases considered above. These frequencies are
calculated as, for the (a), (b), (c), and (d) cases above,
$0.4834$, $0.2441$, $0.3467$, and $0.3418$, (in units of
$\omega_0^{-1}$) respectively. Here the time is measured in units
of $\omega ^{-1}=1$ and hence the trap frequency corresponding to
(\ref{c}) is $\nu_0 = 1/(2\pi)$. The frequency of oscillation was
two times the existing harmonic oscillator frequency for
nonlinearity ${\cal N}=0$. As the frequencies are increased and
reduced by $\sqrt 2$ in cases (a) and (b), two times the existing
frequency for (a) is 2$\nu= \sqrt 2/\pi \simeq 0.45$ and for (b)
is 2$\nu =1/(\sqrt 2 \pi )\simeq 0.23 $. In case of (c) and (d)
the frequency is unchanged and $2\nu = 1/\pi\simeq 0.32$. These
numbers compare well with the respective above results.

We also found that these frequencies are in good agreement with
those calculated using finite difference
scheme\cite{ska_pm_jpb:1}. There are differences when comparing
these frequencies with the corresponding harmonic trap
frequencies, the reason for which is due to the presence of
nonlinearity in the system. We study the frequency of
oscillation as a function of nonlinearity in the following.

\subsubsection{Frequency of oscillation of the condensate}

Having studied various time independent as well as time dependent
problems, we further investigate the frequency of oscillation of
the condensate due to a variation of nonlinearity. When the
nonlinear term is added in sufficiently large steps during the
preparation of the condensate numerically, a breathing type
oscillation with some frequency, $\nu$ occurs which is determined
solely by the final nonlinearity. This frequency can be extracted
by exploiting the Fourier transform on the time series of the
\begin{figure}[!ht]
\centering\includegraphics[width=0.95\columnwidth]{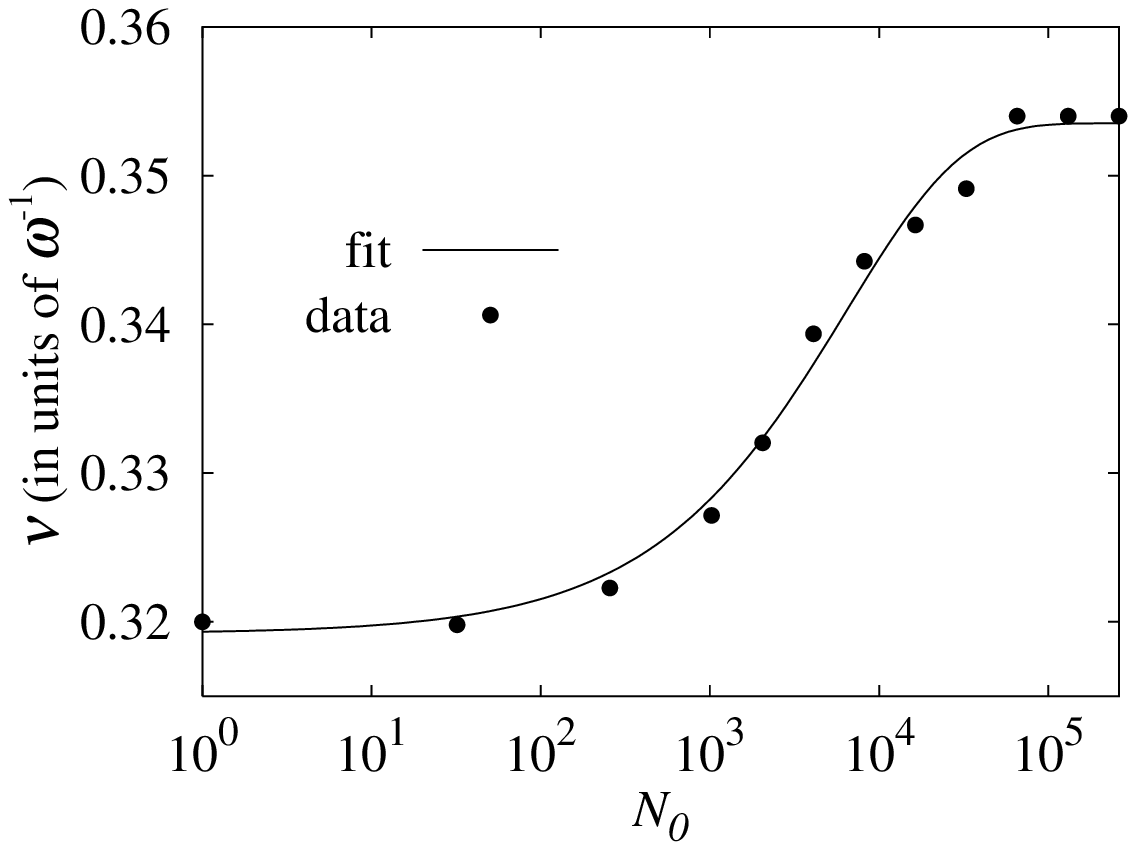}
\caption{Plot showing frequency of oscillation versus the $N_0$}
\label{f2}
\end{figure}
mean distance, $\langle r \rangle_{\mbox{rms}}$ from numerical
simulation. We express the final nonlinearity in terms of the
final value of $N_0$. We found that the frequency of oscillation
increases as the final number of atoms $N_0$ is increased and it
saturate to a final value $0.354$ when $N_0 > 2^{16} \simeq 7\times 10^{4}$.
Figure \ref{f2} shows the of frequency oscillation $\nu$ as a
function of nonlinearity in the condensate. By fitting the
numerical data for the frequencies, the frequency of oscillation
as a function of final $N_0$ can be represented by the following
relation,
\begin{equation}
\nu(N_0) = \nu_0' + \frac{1}{b}\exp\left(-\frac{1}{c}N_0^\gamma\right).
\label{eq:fit}
\end{equation}
The parameters are calculated as $\nu_0' = 0.3536$, $b=-28.6858$,
$c=227.26$ and $\gamma=0.6202$.

\section{Summary and Conclusions}
\label{conclude}

We propose the use of pseudospectral method to solve time
dependent nonlinear Gross-Pitaevskii equation describing
Bose-Einstein condensation in spherically symmetric trap
potential. In particular, we emphasize on the advantages of using
Laguerre polynomial interpolation for the spatial differentiation
operator and we study both stationary as well as time evolution
problems. In the stationary problem, the radial wavefunction and
the ground state energy are calculated by solving the GP equation
for different nonlinearity strengths which determines the number
of atoms in the condensate. We noted that the energy values are
calculated more accurately with very few number of discretization
points (nodes). For the time evolution problem, first we study the
effect of sudden changes in the interatomic scattering length or
harmonic oscillator trap potential in the condensate. We have also
made an analysis on the variation of frequency of oscillation as a
function of the strength of nonlinearity.

This paper summarizes our attempts to understand certain dynamical
aspects of Bose-Einstein condensation as a nonlinear dynamical
system. We found that the present approach based on pseudospectral
differentiation matrices for solving Gross-Pitaevskii equation is
more effective when compared to finite-difference schemes. Though,
we are solving the problem effectively in one-dimension because of
the spherically symmetric trap, the anisotropic problems can be
easily tackled with similar approach\cite{ska_pm_jpb:2}.

\section*{Acknowledgments}
The work of PM is supported by DST, Govt. of India and SKA acknowledges 
CNPq of Brasil

\end{document}